# Machine Learning Framework for RF-Based Drone Detection and Identification System

Olusiji O. Medaiyese, Abbas Syed, and Adrian P. Lauf, *Member, IEEE*

*Abstract*—The emergence of drones has added new dimension to privacy and security issues. There are little or no strict regulations on the people that can purchase or own a drone. For this reason, people can take advantage of these aircraft to intrude into restricted or private areas. A Drone Detection and Identification (DDI) system is one of the ways of detecting and identifying the presence of a drone in an area. DDI systems can employ different sensing technique such radio frequency (RF) signals, video, sounds and thermal for detecting an intruding drone. In this work, we propose a machine learning RF-based DDI system that uses low band RF signals from drone-to-flight controller communication. We develop three machine learning models using the XGBoost algorithm to detect and identify the presence of a drone, the type of drones and the operational mode of drones. For these three XGBoost models, we evaluated the models using 10-fold cross validation and we achieve average accuracy of 99.96%, 90.73% and 70.09% respectively.

## I. INTRODUCTION

In our world today, we continue to see great increases in the application of robots (airborne or non-airborne robots) in different aspects of our lives. Drones are typical examples of airborne robots which are unmanned vehicles. The application of drones is not limited to military use as there is a continuous usage in civilian sector. To name a few, some of these applications range from logistical operations, reconnaissance, search-and-rescue, disaster assessment and so on [1], [2]. According to the Federal Aviation Administration (FAA), there are more than 1.5 million drones registered in the United States. Twenty-eight percent of these registered drones are used for commercial purposes while the remaining percentage are acquired for recreational use [3].

Because benign applications of drones are numerous and no strict regulations exist on who can buy and operate a drone, there is also a rise in safety and security concerns. One common security concern is the use of the drone to invade secured or restricted area. Drones can also be utilized by extremist or terrorist groups to carry explosive payloads or chemicals to a targeted vicinity which could endanger public safety [4].

O. O. Medaiyese is with the Computer Science and Engineering Department, University of Louisville, Louisville, KY 40292 USA (phone: 5024729914; e-mail: o0meda01@louisville.edu).
A. Syed is with the Computer Science and Engineering Department, University of Louisville, Louisville, KY 40292 USA (e-mail: edum0syed03@louisville.edu).
Adrian. P. Lauf is with the Computer Science and Engineering Department, University of Louisville, Louisville, KY 40292 USA (e-mail: aplauf01@louisville.edu).

The use of radar has been frequently adopted in detecting drones or small aircraft [5]. However, there are limitations. For instance, radar cannot distinguish between birds and drones[6]. Several other detection methods have been exploited for use in DDI systems that are based on sensing mechanisms such as sound, video, thermal and radio frequency(RF) [7][8]. In adopting any of these approaches, there are various tradeoffs and these tradeoffs influence the performance of the system. The thermal approach can be severely affected by weather conditions. The effectiveness and efficiency of a sound detection system is inhibited when there is high ambient noise [5] [8]. Similarly, low light visibility and low coverage are setbacks for the use of video detection mechanisms [5].

RF signals emanating from drone can be exploited for DDI. The rotating propellers, communication between the drone and its flight controller, and the vibrating body of the drone are sources of RF signals [9]. Exploiting the statistical distribution of an RF signal by extracting the spectrum features can act as the RF signature for a drone and this is effective irrespective whether or not the environment of propagation for the RF signal is an urban area [8], [9].

Based on the above, we adopted RF signature for DDI system, and the contributions of this work include:

- We have demonstrated that the accuracy of an RF-based drone detection and identification system can be improved using the Extreme Gradient Boosting (XGBoost) algorithm. This implies we do not need computationally expensive algorithms such Deep Neural Network to achieve a better accuracy.

- We propose that minimum feature sets from either the low band or high band of the RF signal is well suitable for building an ML-based model for detecting and identifying drones. However, the low band of RF signatures contains more feature representations useful for input feature vector of our model.

The remainder of this paper is organized as follows: Section II provides an overview of the related work. We define the problem we are solving in Section III. Section IV highlights the ML algorithm used in this work. The description of the dataset, modelling and evaluation of results are provided in Section V. Section VI provides the conclusion and future work.

## II. RELATED WORK

Different approaches have been proposed in the literature for drone detection and identification (DDI) systems. At the core is the integration of artificial intelligence. In [6], a system for detection and identification that employed machine learning for object detection was developed. It uses a surveillance drone which has a mounted camera to capture the video of the environment. If there exists an object like a drone in any frame of the video, it marks the object and image processing procedure will be carried on the marked object. The processed image is fed into an ML model as a feature vector to classify the object as detected. However, an accuracy of 89 percent was achieved. Low light visibility could hinder the performance of the solution.

Audio signatures have also been exploited for DDI. In [4], the authors used the sound from a drone's propeller and motor to develop a sound based DDI using an ML framework which works in a noisy environment. A Support Vector Machine (SVM) classifier was used as the machine learning framework to distinguish flying object based on feature characteristics exhibited by their sounds. The key advantage of their work is that minimal data is required for training the classifier. However, the work was limited to detecting flying object (birds, drones, thunderstorm and airplane) based on sounds. Using sound sensing approach will not perform well when there is high environmental noise [5].

The authors in [9] proposed two approaches for a DDI system, namely active and passive methods. The active method detects a drone by observing the reflected RF signal from the aircraft. In the active method, a transmitter propagates an RF signal toward the drones causing the signal to reflect. The reflected signal can be caused by the drone propellers or the vibrating body of the drone. The reflected signal is then captured by a receiver as a signature for the drone. However, the downside of this approach is that the reflecting signal depends on the orientation of drone, and the distance between the drone and receiver. It was also observed that the signal is not always fully reflected [9]. More so, the changes in environment can cause even further distortion on the reflected signal. This implies that the reflected signal (signature) from a drone may vary with environmental changes. The passive approach involves eavesdropping on the communication signal between the drone and its flight controller because there is a frequent exchange of information between the two devices. However, the signature (RF signal) from eavesdropping on the communication between drone and flight controller depends on the distance between the drone and the DDI system. Increasing the gain of the receiver antenna can help to augment the sensitivity of the RF signal dependence on the distance between the drone and DDI system [9].

Several works have incorporated deep learning, deep belief networks, and other machine learning algorithms such as SVM in DDI systems. For instance, the authors in [10] proposed an RF-based DDI system where deep learning was incorporated as the classification algorithm for detecting and identifying an intruding drones. The proposed algorithm was not only able to identify the presence of drones or type of drones, but it could also detect the operation modes of the drones (hovering, flying, power on/off, and video recording). However, a large dimension of input feature vector which is a sample from the RF signature's spectrum (lower and upper band) was fed into deep neural network (DNN) for classification purpose. The average accuracy results of the classifiers were 99.70%, 84.5% and 46.8% respectively. Based on this work, we proposed XGBoost models that can improve the accuracy of the DDI using half of the feature vector used in [10].

## III. PROBLEM DEFINITION

Predicting a finite or discrete number of variables is a classification problem under supervised learning in ML. Let $y \in D$, given that D is a finite variable which is an identified or registered number of drones permitted to fly or operate in a restricted area and y depends on $\vec{x}$, given that $\vec{x}$ is the feature vector. The feature vector is from the spectrum band of an RF signal from the communication link between the drone and flight controller. The RF signal serves as the RF signature for the drone. Identifying drone $y_i$ is a function of $\vec{x_i}$. Hence $y = F(x)$ is the learning function. To approximate the function $F(x)$, we train XGBoost algorithm using training data T, given that

$$T = \{(\vec{x}_1, y_1), (\vec{x}_2, y_2), (\vec{x}_3, y_3), \ldots, (\vec{x}_D, y_D)\} \quad (1)$$

Because the XGBoost algorithm is decision tree-based, the learning function is represented as a tree rule instead of a mathematical function. The learning function is determined using T, by defining an objective function (both loss function and regularization function) and optimizing the objective function [11].

## IV. XGBOOST ALGORITHM

The motivation for using XGBoost is that it uses fewer resources when compared with deep neural network. XGBoost is a '*scalable machine learning system for tree boosting*"[11]. XGBoost is a form of gradient tree boosting or a gradient boosting machine (GBM) but implemented with a better optimization strategy to improve execution speed and performance[11]. A gradient boosting algorithm is a sequential combination of several predictors (i.e., models) where each predictor is an improvement on its predecessor (e.g., by fitting to the errors of its predecessor) [12]. It is a state-of-the-art ML algorithm and its impact has been seen across the field of data science especially in Kaggle competition [11]. More detailed information about XGBoost can be found in [11]

## V. EXPERIMENT AND ANALYSIS

### A. Data and Sample Description

The publicly available DroneRF dataset was used in this work [13]. Three drones from two different manufacturers were used for the collection of data as detailed in [10], [13]. These are Parrot Bebop, Parrot AR Drone, and DJI Phantom 3.

### B. RF Signature collection and Preprocessing

As mentioned before, we use the DroneRF dataset in this work. The DroneRF dataset contains 227 segments of recorded RF signal strength data in experiments conducted with the 3 drones (i.e., Parrot Bebop, Parrot AR Drone, and DJI Phantom 3). The records contain 10.25 seconds of data for background RF activities (drones off) and approximately 5.25 seconds of drone communication RF data. In addition to the off state, four different modes or states were recorded: *on and connected*, *autonomous hovering*, *flying without any recording* and *flying while recording videos*. The drones were controlled using their respective flight controllers. The RF signal strength data was collected by using two RF signal receivers (NI-USRP 2943R RF receivers) connected to laptops with one measuring strength of lower band (first 40 MHz) of the 80 MHz Wi-Fi channel spectrum and the other receiver measuring the higher band (i.e., the other 40 MHz). Therefore, each segment consists of RF signal strength measurements for the lower band (LB) as well as the upper band (UB). Figure 1 shows the RF signal of the background when no drone is present. Similarly, Figure 2-4 show low band RF signal for the respective drones when the drones are in "*on and connected*" mode. Because each segment consists of a million samples of signal strength measurements, it is not suitable to pass the signals in their original form to the ML algorithm. In order to extract features from the RF signal strength measurements from drone experiments to pass on to the learning algorithm, the Discrete Fourier Transform (DFT) on the recorded RF signal segments (Eq. 2 and Eq. 3) and the magnitude spectrum was extracted which serve as the RF signatures for the drone. These RF signatures are the input feature vector for ML model.

$$X_j^L[k] = \sum_{n=0}^{N-1} x_j^L[n] e^{-j\frac{2\pi nk}{N}} \quad (2)$$

$$X_j^H[k] = \sum_{n=0}^{N-1} x_j^H[n] e^{-j\frac{2\pi nk}{N}} \quad (3)$$

where $X_j^L[k]$ and $X_j^H[k]$ are the N point DFTs of the two parts (LB and UB) of the j[th] segment $x_j^L[n]$ and $x_j^H[n]$. In [10], N was selected to be 2048 and after the authors computed the DFT, they extracted one-sided magnitude spectrum from the DFT output to determine the power distribution in each frequency bin before concatenating the two magnitude spectrums to construct a vector which represents the power distribution over the complete Wi-Fi channel bandwidth. The concatenation was performed by scaling the UB magnitude spectrum and appending it to the end of the LB as

$$X_j = \begin{bmatrix} \|X_j^L\|_{One-sided} & S_j * \|X_j^H\|_{One-sided} \end{bmatrix} \quad (4)$$

where S is the scaling factor which was used to prevent the concatenation operation from introducing discontinuities in the combined magnitude spectrum [10]. The concatenation resulted to 2048 feature vectors which was fed into a DNN for classification. Our work proposes that the either the LB or the UB feature vectors are enough to carry out the classification using the XGBoost algorithm, therefore, reducing the size of the input feature vector from 2048 to 1024.

We experimented using the LB only, UB only, and the combination of LB and UB as input features for the XGBoost model under the same scenarios mentioned in [10]. These scenarios are listed below

i. Presence of drones
ii. Presence of drones and type
iii. Presence of drones, type and operational mode

TABLE I gives a quick summary of the DroneRF dataset. In Case I, there are two classes. This involves detecting the presence of any type of drones in an area. Case II has four classes that involves detecting and identifying the presence of a given type of drone based on the RF signature. Similarly, Case III has ten classes which deals with operational mode of a particular type of drone in a given environment. Mode 1 to 4 represents *On and connected to a flight controller*, *hovering*, *flying without video recording*, and *flying with video recording*, respectively.

TABLE I:
SUMMARY OF DRONERF DATASET [10, Tab 4]

| Case | Class | Segments | Samples ($\times 10^6$) | Sample Percentage (%) |
|---|---|---|---|---|
| [1] I | Drone | 186 | 3720 | 81.94 |
| | No Drone | 41 | 820 | 18.06 |
| [2] II | Bebop | 84 | 1680 | 37 |
| | AR | 81 | 1620 | 35.68 |
| | Phantom | 21 | 420 | 9.25 |
| | No Drone | 41 | 820 | 18.06 |
| [3] III | Bebop mode 1 | 21 | 420 | 9.25 |
| | Bebop mode 2 | 21 | 420 | 9.25 |
| | Bebop mode 3 | 21 | 420 | 9.25 |
| | Bebop mode 4 | 21 | 420 | 9.25 |
| | AR mode 1 | 21 | 420 | 9.25 |
| | AR mode 2 | 21 | 420 | 9.25 |
| | AR mode 3 | 21 | 420 | 9.25 |
| | AR mode 4 | 18 | 360 | 7.93 |
| | Phantom mode 1 | 21 | 420 | 9.25 |
| | No Drone | 41 | 820 | 18.06 |

*C. Performance Evaluation*

We adopted a cross validation (CV) approach that is stratified K-folds to evaluate our models based on certain classification metrics that will be discussed shortly. The K-folds cross validation is a form of out-of-sample testing which involves splitting or dividing the dataset into K folds, where *K-1* folds will used as the training data and one-fold is left as testing data [14]. This process is iteratively done K times to ensure that each fold is used as a testing data. The baseline model for comparison is the DNN model developed by the authors in [10] where 10-fold cross validation was used. Hence, we also use 10-fold cross validation for all our proposed models for this work.

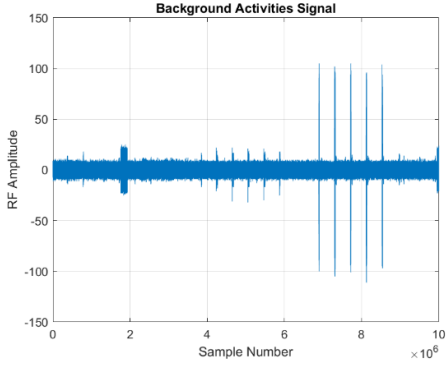

Figure 1. Background RF signal without drone

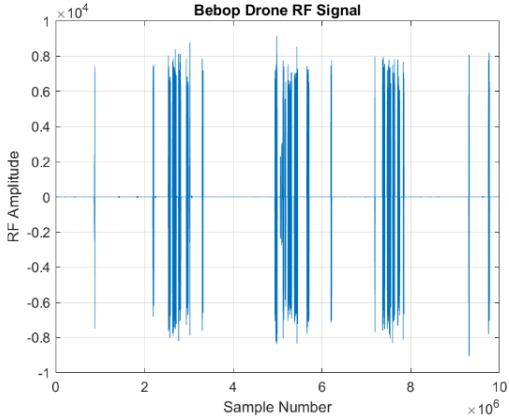

Figure 2. Parrot Bebop RF Signal at "On and Connected"

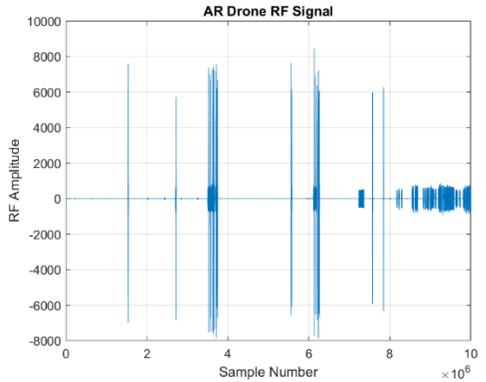

Figure 3. Parrot AR Drone RF Signal at "On and Connected"

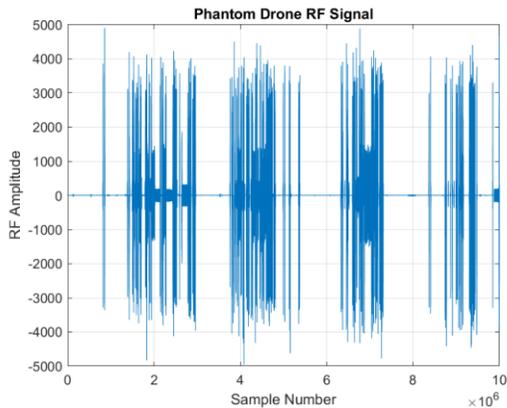

Figure 4. DJI Phantom 3 RF Signal at "On and Connected"

### D. Evaluation Metrics:

There are several metrics for evaluating a classification model. These include accuracy, precision, recall, specificity, F1-score, normalized gini coefficient, Cohen's Kappa, ROC AUC, confusion matrix, and so on. We select the four most common metrics, which are accuracy, precision, recall, and F1-Score.

### E. Result and Analysis

The results of the models are explained based on the three cases listed previously.

#### 1) Detection of the presence of a drone

The average accuracy and the statistical dispersal of the accuracy based on standard deviation from the 10-fold CV using only the lower band of the RF signature as input features are 99.96% and 0.05% respectively. The average F1-score of using just the lower band is approximately equal to 100%. In contrast to using only the upper band of the RF signatures, the accuracy decreases to 98.44% and the standard deviation increases to 2.60%. We used statistical t-tests to compare the mean difference between the accuracy score from the 10-fold CV. Our null hypothesis is that the mean difference of the accuracy scores is equal to zero.

The t-tests in TABLE II shows that there is no significant difference in the mean of the accuracy using either bands for prediction. This shows that the lower band and upper band individually have similar information in detecting the presence of drone in a vicinity.

More so, the t-test when only the lower band and the combination of both the lower and upper band are used as feature vector also revealed no significant difference in the average accuracy. The result of this t-test is shown in

TABLE III

#### 2) Detection of the presence of a drone and type of drone

The lower band of the RF signatures has significant information in detecting and identifying the presence of any type of drone. We achieved an average accuracy of 90.73% with a standard deviation of 9.28% from the CV. The average accuracy decreases using only the upper band as input feature vector to 85.38%, while the spread of the mean increases to 12.31%. This shows a wider variation. However, the statistical test (t-test) depicts that there is no significant difference in the average accuracy when either lower or upper band is used as input feature vector. Similarly, the t-test also shows that there is no difference between the accuracy when we use only lower band and the concatenation of both the upper and lower band.

#### 3) Detection of presence of drone, type of drone and operational mode

Beyond determining the type of drone using the RF signature, the lower band contains more information in its feature space to identify and detect the kind of operations (e.g., flying, hovering, videoing, etc.) performed by a type of drone compared to the upper band. The average accuracy and standard deviation from the CV are 70.09% and 0.89% respectively. However, when only the upper band was used, the mean of accuracy decreased by 23.61% and standard deviation increased to 1.33%. From the statistical test results shown in TABLE II, the null hypothesis was rejected which implies that there is significant difference between the average accuracy of using only lower band and only upper as feature vector.

Furthermore, when we concatenate both the upper and lower band of the RF signatures, we achieved an average accuracy of 70.67% with a standard deviation of 0.67%. Comparing this result with using only the lower band of the RF signals as input features, the t-test results in TABLE III shows that there is no significant difference in the accuracy of the classification.

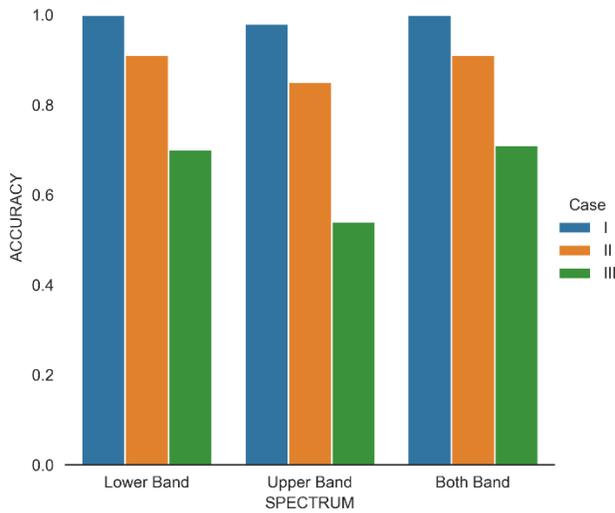

Figure 5. Comparing the average accuracy using lower, upper and both band as input feature vectors for the 3 scenarios

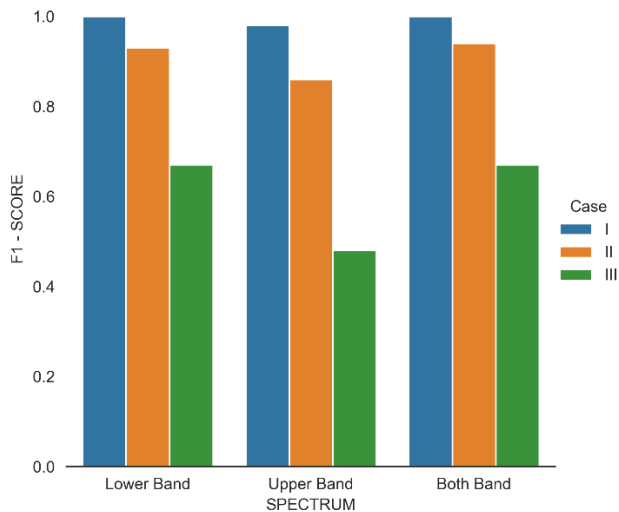

Figure 6. Comparing the average F1-Score using lower, upper and both band as input feature vectors for the 3 scenarios

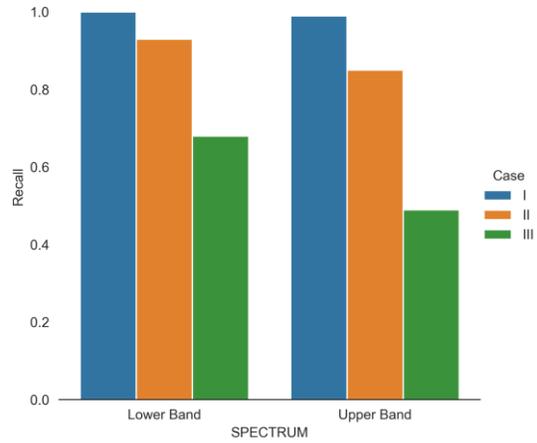

Figure 7. Comparing the average recall using lower and upper band as input feature vectors for the 3 scenarios

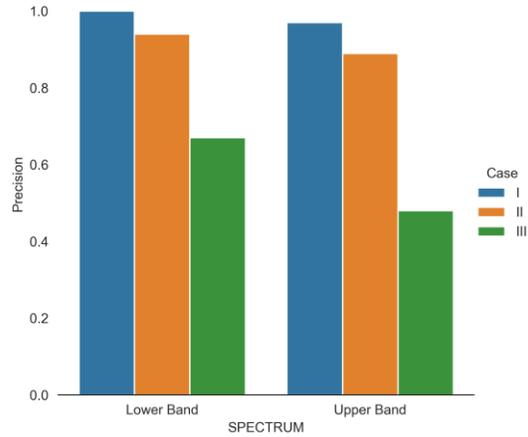

Figure 8. Comparing the average precision using lower and upper band as input feature vectors for the 3 scenarios

In summary, Figure 5 and Figure 6 show the bar charts for the average accuracy and F1 - score respectively when only lower band, only upper band and both bands are used as input feature vectors for DDI system based on the three scenarios considered in this work. Similarly, Figure 7 and Figure 8 show the average recall and precision using the lower and upper band respectively. There was a significant drop in both recall and precision when the upper band RF signal was used for detecting the operational mode of drones as compared to using the lower band.

TABLE II:
PAIRED T-TEST BETWEEN ACCURACY OF THE LOWER BAND AND UPPER BAND SPECTRUM FROM 10-FOLD CROSS VALIDATION TEST

| Case | Null Hypothesis $\mu_d = 0$ at $\alpha = 5\%$ | Confidence interval |
|---|---|---|
| I | Not Rejected | $-0.0037 < \mu_d < 0.0357$ |
| II | Not Rejected | $-0.056 < \mu_d < 0.162$ |
| III | Rejected | $0.1545 < \mu d < 0.1775$ |

TABLE III:
PAIRED T-TEST BETWEEN ACCURACY OF THE LOWER BAND AND BOTH BAND SPECTRUM FROM 10-FOLD CROSS VALIDATION TEST

| Case | Null Hypothesis $\mu_d = 0$ at $\alpha = 5\%$ | Confidence interval |
|---|---|---|
| I | Not Rejected | $-0.0004 < \mu_d < 0.0004$ |
| II | Not Rejected | $-0.0924 < \mu_d < 0.0904$ |
| III | Not Rejected | $-0.0138 < \mu_d < 0.0018$ |

From the above discussion of the results, we have been able to justify that capturing the lower band of RF signal as the RF signature is suitable for detecting and identifying drones. We also compared the accuracy of using our proposed approach to the recently-proposed DNN algorithm by the authors in [10] where all the spectrum (both the lower and upper band) of the RF signatures were used as the feature vectors.

In TABLE IV, we compare our results with the results that were acquired using a DNN algorithm in [10]. It can be observed that there are significant improvements using XGBoost algorithm and only the lower band of the RF signature as input feature vector. In Case I, we managed to increase the accuracy by 0.26%. Similarly, there was significant increase in the Case II and III where the accuracies were increased by 7.37% and 49.76% respectively.

TABLE IV:
PAIRED T-TEST BETWEEN ACCURACY OF THE LOWER BAND AND BOTH BAND SPECTRUM FROM 10-FOLD CROSS VALIDATION TEST

| Case | Average Accuracy | |
|---|---|---|
|  | XGBoost | DNN [10] |
| I | 99.96% | 99.70% |
| II | 90.73% | 84.5% |
| III | 70.09% | 46.8% |

## VI. CONCLUSION AND FUTURE WORK

As demonstrated in this work that XGBoost algorithm performs well in the domain of DDI using RF-based signatures. No wonder it is a wining ML algorithm in many industry-based problems. Based on our results, we concluded that the capturing the lower band spectrum of the RF signal from the communication between a drone and its flight controller is sufficient in ML based DDI systems as input feature vectors. Although the upper band spectrum has some information that can be exploited for DDI systems, the lower band contains more information that can be used for drone detection and identification. The accuracy of our classifier decreases when using RF signature to detect and identify operational modes of drones. This implies that the effectiveness of using frequency analysis of the RF signature to detect activities performed by drones is low.

While the dataset used for this work was acquired under a controlled environment where there is little or no interference from other wireless devices, we plan to exploit how interference from both static and dynamic wireless devices can affect the feature property of RF-signature spectrum in future work. It would also be pertinent to look into other feature extraction schemes to characterize drone activities.

REFERENCES

[1] M. Erdelj, E. Natalizio, K. R. Chowdhury, and I. F. Akyildiz, "Help from the Sky: Leveraging UAVs for Disaster Management," in IEEE Pervasive Computing, 2017, vol. 16, no. 1, pp. 24–32.

[2] A. C. Watts, V. G. Ambrosia, and E. A. Hinkley, "Unmanned Aircraft Systems in Remote Sensing and Scientific Research: Classification and Considerations of Use," Remote Sens., vol. 4, pp. 1671–1692, 2012, doi: 10.3390/rs4061671.

[3] "UAS by the Numbers," Federal Aviation Administration. [Online]. Available: https://www.faa.gov/uas/resources/by_the_numbers/. [Accessed: 02-Feb-2020].

[4] M. Z. Anwar, Z. Kaleem, and A. Jamalipour, "Machine Learning Inspired Sound-Based Amateur Drone Detection for Public Safety Applications," in IEEE Transactions on Vehicular Technology, 2019, vol. 68, no. 3, pp. 2526–2534.

[5] P. Nguyen, H. Truong, M. Ravindranathan, A. Nguyen, R. H. Han, and T. Vu, "Cost effective and passive RF-based Drone presence detection and characterization," in GetMobile: Mobile Computing and Communications, 2017, vol. 21, no. 4, pp. 30–34.

[6] D. Lee, W. G. La, and H. Kim, "Drone Detection and Identification System using Artificial Intelligence," in 2018 International Conference on Information and Communication Technology Convergence (ICTC), 2018, pp. 1131–1133, doi: 10.1109/ICTC.2018.8539442.

[7] G. Ding, Q. Wu, L. Zhang, Y. Lin, T. A. Tsiftsis, and Y. Yao, "An Amateur Drone Surveillance System Based on the Cognitive Internet of Things," IEEE Commun. Mag., vol. 56, no. 1, pp. 29–35, 2018.

[8] X. Shi, C. Yang, W. Xie, C. Liang, Z. Shi, and J. Chen, "Anti-Drone System with Multiple Surveillance Technologies: Architecture, Implementation, and Challenges," IEEE Commun. Mag., vol. 56, no. 4, pp. 68–74, 2018.

[9] P. Nguyen, M. Ravindranathan, A. Nguyen, R. Han, and T. Vu, "Investigating Cost-effective RF-based Detection of Drones," in DroNet '16: Proceedings of the 2nd Workshop on Micro Aerial Vehicle Networks, Systems, and Applications for Civilian Use, 2016, pp. 17–22, doi: https://doi.org/10.1145/2935620.2935632.

[10] M. F. Al-sa, A. Al-ali, A. Mohamed, and T. Khattab, "RF-based drone detection and identification using deep learning approaches: An initiative towards a large open source drone database," Futur. Gener. Comput. Syst., vol. 100, pp. 86–97, 2019, doi: 10.1016/j.future.2019.05.007.

[11] T. Chen and C. Guestrin, "XGBoost: A Scalable Tree Boosting System," in KDD '16: Proceedings of the 22nd ACM SIGKDD International Conference on Knowledge Discovery and Data Mining, 2016, pp. 785–794.

[12] A. Géron, Hands-On Machine Learning with Scikit-Learn and TensorFlow, 1st Editio. O'Reilly Media, 2017.

[13] M. S. Allahham, M. F. Al-Sa'd, A. Al-Ali, A. Mohamed, T. Khattab, and A. Erbad, "DroneRF dataset: A dataset of drones for RF-based detection, classification and identification," Data Br., vol. 26, 2019, doi: 10.1016/j.dib.2019.104313.

[14] T. Wong, "Performance evaluation of classification algorithms by k-fold and leave-one-out cross validation," Pattern Recognit., vol. 48, no. 9, pp. 2839–2846, 2015, doi: 10.1016/j.patcog.2015.03.009.